\documentclass[a4paper,10pt]{article}
\usepackage[utf8]{inputenc}
\usepackage[T1]{fontenc}
\usepackage{microtype}
\usepackage[italian]{babel}
\usepackage{amsmath}
\usepackage{amsthm}
\usepackage{algorithmic} 
\usepackage{algorithm}
\usepackage{amsfonts}
\usepackage{graphicx}
\usepackage{booktabs}
\usepackage{subfigure}
\usepackage{tikz}

\usepackage{authblk}
\usepackage{resizegather}
\usepackage{wasysym}
\usepackage{textcomp}
\usepackage{tikz}
\usepackage{url}
\usepackage{fancyhdr, lipsum}

\title{FIEMS\footnote{Fast Italian Energy Market Simulator} : algortimo di simulazione del mercato elettrico italiano.}
\author{Matteo Gardini \and Marco Diana}
\date{\today}
\begin{document}

\maketitle

\begin{abstract}
\noindent Il documento illustra l'algoritmo del mercato elettrico utilizzato in Italia a partire dalla liberalizzazione del mercato elettrico avvenuta nel 2004. Vengono poi forniti i dettagli di implementazione in Matlab di una sua versione semplificata, capace di produrre risultati accettabili in un tempo estremamente breve. 
\end{abstract}

\section{Introduzione}
\begin{figure}[!h]
	\centering
	\includegraphics[scale=0.7]{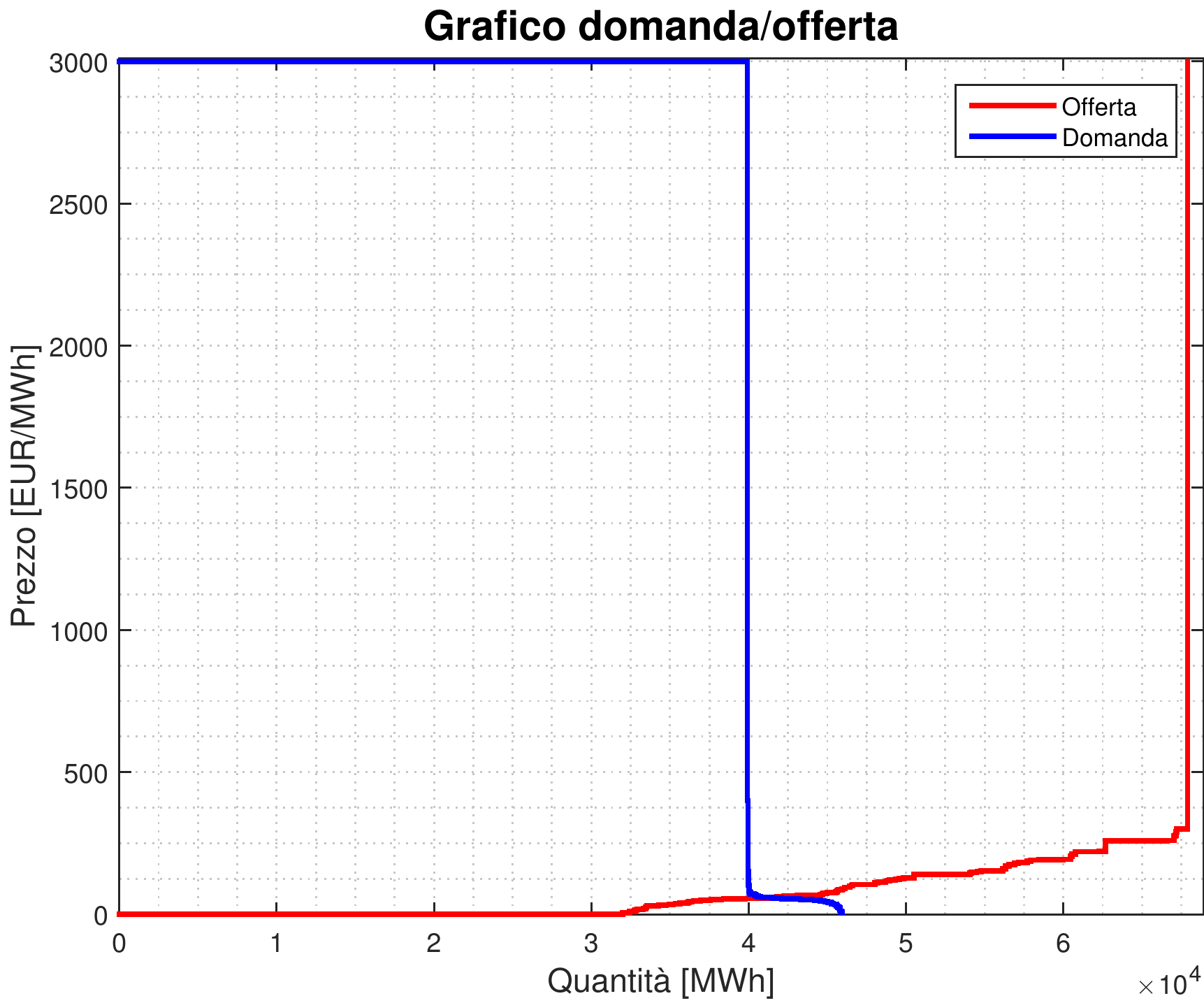}
	\caption{Curva della domanda e dell'offerta per il sistema Italia completo}
	\label{Curva_Domanda_Offerta}
\end{figure}

Con la delibera 111/06 del 1999, nel 2004 è avvenuta la liberalizzazione del mercato dell'energia elettrica ed del gas. In seguito si è visto fiorire un notevo numero di operatori che, giorno per giorno, cercano di garantirsi l'energia al prezzo più vantaggioso possibile. Responsabile della gestione delle offerte nel mercato del giorno prima (MGP) è il gestore del mercato energetico (GME) che, ogni giorno per ogni ora, determina il prezzo dell'energia elettrica sulla base delle offerte di acquisto e vendita presentate dai vari operatori di mercato. La notevole complessità di questo sistema può essere sintetizzata nel seguente modo. Ogni operatore presenta un'offerta di vendita costituita da una coppia quantità/prezzo che indica che l'operatore è disposto a vendere al massimo l'ammontare di energia specificato dalla quantità ad un prezzo che sia non inferiore a quello indicato. Contrariamente un'operatore che presenta un'offerta di acquisto è disposto ad acquistare al massimo l'ammontare di energia indicato dalla quantità ad un prezzo che non sia superiore a quello indicato. Il GME raccoglie tutte le offerte, dettagliate per ora, determina il prezzo e decide quali offerte vengono accettare e quali invece sono rigettate, tenendo conto che l'energia elettrica non può essere immagazzinata e quindi la quantità venduta deve essere esattamente pari alla quantità acquistata. Quello che il GME fa non è altro che impilare per prezzo crescente le offerte di vendita e per prezzo decrescente le offerte di acquisto, incrociare la curva della domanda e dell'offerta e determinare il prezzo dell'energia per quell'ora come mostrato in Figura $\ref{Curva_Domanda_Offerta}$. 

Il procedimento per la determinazione del prezzo, se fosse così, sarebbe estremamente elementare. Quello che complica notevolmente la situazione è che l'energia elettrica viaggia su connettori di capacità fisica finita. Per questo motivo, nella risoluzione del problema determinazione del prezzo dell'energia, bisogna essere certi di non violare i limiti fisici imposti dal sistema. Ne risulta che incrociare la curva della domanda e dell'offerta non è sufficiente a determinare il prezzo dell'energia e a rispettare i vincoli. Il problema che va risolto è (nel caso più semplice) un problema di programmazione lineare vincolata. Una volta risolto questo problema di ottimizzazione, risulta che il benessere del sistema è il massimo possibile.\\
Nel seguente articolo illustreremo un algoritmo \textquotedblleft semplificato\textquotedblright\ che permette di risolvere il problema. Ne forniremo una descrizione da un punto di vista matematico, per poi passare a fornirne i dettagli implementativi in Matlab e, da ultimo, ne analizzeremo la bontà rispetto ai prezzi e ai transiti del mercato reale.

\section{Formulazione Matematica}
L'obiettivo di determinare i prezzi zonali in Italia sulla base delle quantità di energia in vendita (OFF) e in acquisto (BID) da un punto di vista matematico si riduce, nel caso più semplice, alla risoluzione di un problema di ottimizzazione lineare vincolata come descritto in \cite{GMEParz2004}. Una trattazione rigorosa e una descrizione dettagliata dell'algoritmo completo può essere trovata in \cite{GMEComplete2004} e \cite{CESI2003}.
In questa sezione presenteremo il problema nella sua formulazione più semplice.

Iniziamo con il definire alcune variabili fondamentali per scrivere il problema in linguaggio matematico.

\begin{itemize}
	\item $N$: numero totale di zone in cui il paese è diviso. Attualmente in italia sono presenti 22 zone.
	\item $K_C$: numero totale di offerte di acquisto di energia.
	\item $K_G$: numero totale di offerte di vendita di energia.
	\item $C_{ij}=C_{ji}$: indice della matrice di adiacenza $C$ rappresentante il grafo della rete elettrica. $C_{ij}=1$ se esiste una connessione tra la zona $i$ e la zona $j$, $0$ altrimenti.
	\item $M$: numero totale linee intra-zonali che vengono monitorate per evitare congestioni. 
	\item $z$: indice che denota una determinata zona. $z=1,\dots,N$.  
	\item $i,j$: indici usati per indicare se tra due zone esiste un transito di energia. $i,j \in \left\{1,2,\dots,N\right\}$
	\item $\alpha$: indice usato per di indicare una trasmissione di energia intra-zonale sulla quale il flusso di energia è limitato per varie ragioni. $\alpha=1,2,\dots,M$.
	\item $k_c$: indice che denota le offerte di acquisto di energia. $k_c=1,\dots,K_C$.
	\item $k_g$: indice che denota le offerte di vendita di energia. $k_g=1,\dots,K_G$.
	\item $PV_{k_g},QOV_{k_g}$: coppia prezzo-quantità di energia in vendita associate all'offerta di vendita $k_g$ per tutti i $k_g=1,\dots,K_G$.
	\item $QVMIN$: minima quantità di energia accettata se un'offerta di generazione è accettata.
	\item $PA_{k_c},QOA_{k_c}$: coppia prezzo-quantità di energia in acquisto associata all'offerta di acquisto $k_c$ per tutti i $k_c=1,\dots,K_C$.
	\item $MAX_{ij}$: massima quantità di energia in transito possibile sull'interconnessione da $i$ a $j$.
	\item $S_{ij}^z$: contributo di un $MW$ di iniezione di energia nella zona $z$ al reale flusso di energia sul transito che connette la zona $i$ e $j$. Questi coefficienti sono calcolati dai coefficienti $C_{ij}$ e riflettono le appropriate impedenze nel caso nella rete fossero presenti dei cicli.
	\item $A_{\alpha}^z$: contributo di un $MW$ di iniezione di energia nella zona $z$ al reale flusso su alcune connessioni intrazonali $\alpha$.
	\item $b_{\alpha}$: massimo valore di energia che può fluire sulla linea di trasmissione $\alpha$ con $\alpha=1,\dots,M$.
	\item $QV_{k_g},QA_{k_c}$: quantità accettata della singola offerta di generazione o cunsumo per ogni $k_c$,$k_g$.
	\item $\rho_z$: prezzo dell'energia nella zona $z$.
	\item $\lambda$: variabile duale associata al vincolo di bilancio di energia.
	\item $\mu_{ij}$: per ogni $i\ne j$ e $i,j\in \left\{1,2,\dots,N\right\}$ tali che $C_{ij}\ne 0$. Variabili duali associati al vincolo di scambio di energia tra zone. Si noti che $\mu_{ij}$ sarà nullo quando il transito tra le zone $i$ e $j$ non sarà saturato.
	\item $\nu_{\alpha}$: per $\alpha=1,\dots,M$ variabili duali associati con il vincolo di trasmissione intra-zonale.
	\item $\rho_{SL}:$ prezzo nazionale in assenza di limiti di transito tra zone. 
		
\end{itemize}

\begin{figure}[!h]
\centering
\includegraphics[scale=0.6]{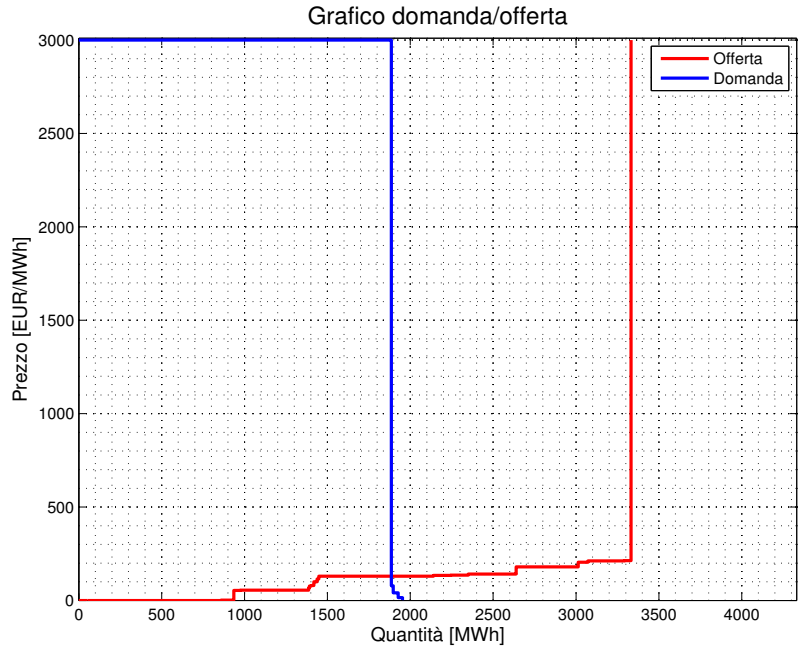}
\caption{Funzione Benessere del sistema}
\label{Fun_Ben_Sistema}
\end{figure}

Il problema di ottimizzazione può essere formulato come mostrato in seguito. E' necessario massimizzare la seguente funzione obiettivo:
\begin{equation}
\max_{QA_{k_c},QV_{k_g}} \left\{\sum_{k_c=1}^{K_C} PA_{k_c}QA_{k_c} - \sum_{k_g=1}^{K_G} PV_{k_g}QV_{k_g}  \right\},
\label{Fun_Obj}
\end{equation}
in cui le variabili decisionali sono $QA_{k_c}$ e $QV_{k_g}$ cioè, rispettivamente, la quantità di energia accettata per ogni singola offerta di vendita $QOV_{kg}$ e di aquisto $QOA_{kc}$. 
Tale funzione rappresenta il benessere del sistema elettrico e corrisponde a massimizzare l'area a sinistra compresa tra le curve in Figura $\eqref{Fun_Ben_Sistema}$

La massimizzazione della funzione obiettivo $\eqref{Fun_Obj}$ è soggetta ai seguenti vincoli lineari.
Innanzi tutto, non è possibile che venga accettata una quantità di energia, sia essa in vendita o in acquisto, superiore alla quantità offerta. Tale richiesta è espressa dai vincoli $\eqref{Con_QA}$ e $\eqref{Con_QV}$, che limitano il campo di scelta delle variabili decisionali.
\begin{equation}
0\le QA_{k_c} \le QOA_{k_c} \; \forall \; k_c\in{1,2,\dots,KC}
\label{Con_QA}
\end{equation}

\begin{equation}
0\le QV_{k_g} \le QOV_{k_g} \; \forall \; k_g\in{1,2,\dots,KG}
\label{Con_QV}
\end{equation}

Condizione fondamentale per il sistema elettrico è che esso risulti bilanciato, ovvero che la quantità di energia immessa dalle unità di produzione sia esattamente l'energia richiesta dalle unità di consumo non essendoci, come noto, possibilità di accumulo di energia elettrica. Questa richiesta è esplicitata dal vincolo $\eqref{Con_BIL}$ che impone che la quantità di energia in vendita sia pari a quella in acquisto.

\begin{equation}
\sum_{k_c=1}^{K_C} QA_{k_c}=\sum_{k_g=1}^{K_G} QV_{k_g}
\label{Con_BIL}
\end{equation}

Le varie macrozone in cui è diviso il mercato elettrico italiano sono interconnesse tra loro da connettori che, per loro natura intrinseca, ammettono un passaggio massimo di energia all'interno dell'ora. Non risulta quindi possibile trasmettere una quantità di energia che sia superiore al massimo consentito. Il vincolo $\eqref{Con_LIMTRANS}$, gestisce questa richiesta, richiedento che il flusso di energia su un determinato connettore tra la zona $i$ e la zona $j$ sia inferiore al valore indicato da $MAXF_{ij}$.

\begin{equation}
\sum_{z=1}^N S_{ij}^{z} \left[ \sum_{k_g \in aggr \, bus \,z} QV_{k_g} - \sum_{k_c \in aggr \, bus \,z} QA_{k_c}  \right] \le MAXF_{ij} \; \forall i,j \in \left\{1,\dots,N\right\}, C_{ij}\ne 0, i\ne j
\label{Con_LIMTRANS}
\end{equation}

Da ultimo, è necessario richiedere che non siano presenti congestioni all'interno della rete, cosa che potrebbe accadere nel momento in cui fossero presenti dei cicli. Tale eventualità è scongiurata dal vincolo in $\eqref{Con_CONG}$. 

\begin{equation}
\sum_{z=1}^{N} A_{\alpha}^{z} \left[ \sum_{k_g \in aggr \, bus \,z} QV_{k_g} - \sum_{k_c \in aggr \, bus \,z} QA_{k_c} \right]\le b_{\alpha} \; \forall \, \alpha=1,2,\dots,M.
\label{Con_CONG}
\end{equation}

Un'osservazione e d'obbligo arrivati a questo punto. Nel caso in cui la rete elettrica godesse della topologia di albero, allora tutti i coefficienti $S_{ij}$ sarebbero unitari, mentre i coefficienti $A_{\alpha}$ risulterebbero nulli. La trattazione relativa al caso generale può essere trovata in. \\
\par Una volta risolto i problema di programmazione lineare il prezzo nelle singole zone può essere calcolato come:
\begin{equation}
\rho_z=\lambda_{eq}-\sum_{ij \; s.t. \; C_{ij}=1} \lambda_{ij}S_{ij}^z - \sum_{\alpha=1,2,\dots,M}\nu_{\alpha} A_{\alpha}^z.
\end{equation}

\section{Implementazione numerica}
Per l'implementazione dell'algoritmo descritto nella sezione precedente si è scelto di ultilizzare il software $Matlab$ versione $2014a$, dotandosi dei pacchetti Optimization Toolbox e Global Optimization Toolbox.\\
Nel seguito descriveremo l'implementazione fatta dell'algoritmo, soffermandoci su criticità ed aspetti cruciali.\\
Ogni esecuzione dell'algoritmo risolve il problema della sezione precedente limitatamente ad una sola ora. Risolvendo quindi il problema di ottimizzazione per tutte le ore della giornata precedente sarà possibile determinare i profilo di prezzi orari per tutte le zone del mercato italiano.

\subsection{Definizione della rete elettrica}
La rete elettrica italiana, che gode della topologia di grafo, è rappresentata all'interno di $Matlab$ sotto forma di matrice di adiacenza $\boldsymbol{G}$ in cui $G_{ij}=1$ se esiste un collegamento tra la zona $i$ e la zona $j$, $0$ altrimenti. Fino al 25 febbraio 2015 la rete italiana ha goduto della topologia di albero:successivamente a questa data in concomitanza con l’avvio dell’applicazione dell’algoritmo di Market-Coupling Europeo da parte di GME e di una gestione migliorata dei collegamenti tra Italia, Corsica e Sardegna, si è venuto a creare un \textquotedbl anello\textquotedbl\ tra le zone  CNOR-CSUD-SARD-CORS-CNOR. La trattazione presentata fa riferimento alla situazione italiana antecedente a questa data anche se esperimenti numerici hanno dimostrato che l'algoritmo produce buoni risultati anche nel caso di topologia a grafo.\\
Le zone di mercato, comprensive quelle introdotte con il Market Coupling, sono ventidue e sono: FRAN, SVIZ, AUST, SLOV,	BSP,	NORD,	CNOR,	SARD,	CORS,	COAC,	CSUD,	SUD,	FOGN,	BRNN,	GREC,	ROSN,	SICI,	PRGP,	MFTV,	XFRA,	XAUS,	MALT.\\
La rete italiana è rappresentata nella figura seguente.

\begin{tikzpicture}

\node[rectangle,fill=gray!20] (n20) at (3,15) {XFRA};
\node[rectangle,fill=gray!20] (n23) at (7,15) {XSVI};
\node[rectangle,fill=gray!20] (n21) at (11,15) {XAUS};
\node[rectangle,fill=gray!20] (n5) at (15,15) {BSP};
\node[rectangle,fill=gray!20] (n1) at (3,13) {FRAN};
\node[rectangle,fill=gray!20] (n2) at (7,13) {SVIZ};
\node[rectangle,fill=gray!20] (n3) at (11,13) {AUST};
\node[rectangle,fill=gray!20] (n4) at (15,13) {SLOV};
\node[rectangle,fill=gray!20] (n6) at (12,11) {NORD};
\node[rectangle,fill=gray!20] (n19) at (15,11) {MFTV};
\node[rectangle,fill=gray!20] (n10) at (4,7) {COAC};
\node[rectangle,fill=gray!20] (n9) at (8,9) {CORS};
\node[rectangle,fill=gray!20] (n7) at (12,9) {CNOR};
\node[rectangle,fill=gray!20] (n8) at (8,7) {SARD};
\node[rectangle,fill=gray!20] (n11) at (12,7) {CSUD};
\node[rectangle,fill=gray!20] (n12) at (12,6) {SUD};
\node[rectangle,fill=gray!20] (n18) at (3,5) {PRGP};
\node[rectangle,fill=gray!20] (n16) at (7,5) {ROSN};
\node[rectangle,fill=gray!20] (n13) at (12,5) {FOGN};
\node[rectangle,fill=gray!20] (n14) at (15,5) {BRNN};
\node[rectangle,fill=gray!20] (n17) at (3,3) {SICI};
\node[rectangle,fill=gray!20] (n22) at (7,3) {MALT};
\node[rectangle,fill=gray!20] (n15) at (15,3) {GREC};

\draw (n20) to  (n1);
\draw (n23) to  (n2);
\draw (n21) to  (n3);
\draw (n5) to  (n4);
\draw (n1) to  (n6);
\draw (n2) to  (n6);
\draw (n3) to  (n6);
\draw (n4) to  (n6);
\draw (n6) to  (n7);
\draw (n7) to  (n9);
\draw (n8) to  (n11);
\draw (n10) to  (n8);
\draw (n8) to  (n9);
\draw (n7) to  (n11);
\draw (n11) to  (n12);
\draw (n12) to  (n16);
\draw (n18) to  (n17);
\draw (n22) to  (n17);
\draw (n15) to  (n14);
\draw (n12) to  (n13);
\draw (n16) to  (n17);
\draw (n14) to  (n12);

\label{rete_italiana}
\end{tikzpicture}

\clearpage
La matrice di adiacenza $\boldsymbol{G}$ utilizzata è la seguente:

\begin{equation}
\boldsymbol{G}=
\begin{bmatrix}
0 & 0 & 0 & 0 & 0 & 1 & 0 & 0 & 0 & 0 & 0 & 0 & 0 & 0 & 0 & 0 & 0 & 0 & 0 & 1 & 0 & 0 & \\ 
0 & 0 & 0 & 0 & 0 & 1 & 0 & 0 & 0 & 0 & 0 & 0 & 0 & 0 & 0 & 0 & 0 & 0 & 0 & 0 & 0 & 0 & \\ 
0 & 0 & 0 & 0 & 0 & 1 & 0 & 0 & 0 & 0 & 0 & 0 & 0 & 0 & 0 & 0 & 0 & 0 & 0 & 0 & 1 & 0 & \\ 
0 & 0 & 0 & 0 & 1 & 1 & 0 & 0 & 0 & 0 & 0 & 0 & 0 & 0 & 0 & 0 & 0 & 0 & 0 & 0 & 0 & 0 & \\ 
0 & 0 & 0 & 1 & 0 & 0 & 0 & 0 & 0 & 0 & 0 & 0 & 0 & 0 & 0 & 0 & 0 & 0 & 0 & 0 & 0 & 0 & \\ 
1 & 1 & 1 & 1 & 0 & 0 & 1 & 0 & 0 & 0 & 0 & 0 & 0 & 0 & 0 & 0 & 0 & 0 & 0 & 0 & 0 & 0 & \\ 
0 & 0 & 0 & 0 & 0 & 1 & 0 & 0 & 1 & 0 & 1 & 0 & 0 & 0 & 0 & 0 & 0 & 0 & 0 & 0 & 0 & 0 & \\ 
0 & 0 & 0 & 0 & 0 & 0 & 0 & 0 & 1 & 1 & 1 & 0 & 0 & 0 & 0 & 0 & 0 & 0 & 0 & 0 & 0 & 0 & \\ 
0 & 0 & 0 & 0 & 0 & 0 & 1 & 1 & 0 & 0 & 0 & 0 & 0 & 0 & 0 & 0 & 0 & 0 & 0 & 0 & 0 & 0 & \\ 
0 & 0 & 0 & 0 & 0 & 0 & 0 & 1 & 0 & 0 & 0 & 0 & 0 & 0 & 0 & 0 & 0 & 0 & 0 & 0 & 0 & 0 & \\ 
0 & 0 & 0 & 0 & 0 & 0 & 1 & 1 & 0 & 0 & 0 & 1 & 0 & 0 & 0 & 0 & 0 & 0 & 0 & 0 & 0 & 0 & \\ 
0 & 0 & 0 & 0 & 0 & 0 & 0 & 0 & 0 & 0 & 1 & 0 & 1 & 1 & 0 & 1 & 0 & 0 & 0 & 0 & 0 & 0 & \\ 
0 & 0 & 0 & 0 & 0 & 0 & 0 & 0 & 0 & 0 & 0 & 1 & 0 & 0 & 0 & 0 & 0 & 0 & 0 & 0 & 0 & 0 & \\ 
0 & 0 & 0 & 0 & 0 & 0 & 0 & 0 & 0 & 0 & 0 & 1 & 0 & 0 & 1 & 0 & 0 & 0 & 0 & 0 & 0 & 0 & \\ 
0 & 0 & 0 & 0 & 0 & 0 & 0 & 0 & 0 & 0 & 0 & 0 & 0 & 1 & 0 & 0 & 0 & 0 & 0 & 0 & 0 & 0 & \\ 
0 & 0 & 0 & 0 & 0 & 0 & 0 & 0 & 0 & 0 & 0 & 1 & 0 & 0 & 0 & 0 & 1 & 0 & 0 & 0 & 0 & 0 & \\ 
0 & 0 & 0 & 0 & 0 & 0 & 0 & 0 & 0 & 0 & 0 & 0 & 0 & 0 & 0 & 1 & 0 & 1 & 0 & 0 & 0 & 1 & \\ 
0 & 0 & 0 & 0 & 0 & 0 & 0 & 0 & 0 & 0 & 0 & 0 & 0 & 0 & 0 & 0 & 1 & 0 & 0 & 0 & 0 & 0 & \\ 
0 & 0 & 0 & 0 & 0 & 0 & 0 & 0 & 0 & 0 & 0 & 0 & 0 & 0 & 0 & 0 & 0 & 0 & 0 & 0 & 0 & 0 & \\ 
1 & 0 & 0 & 0 & 0 & 0 & 0 & 0 & 0 & 0 & 0 & 0 & 0 & 0 & 0 & 0 & 0 & 0 & 0 & 0 & 0 & 0 & \\ 
0 & 0 & 1 & 0 & 0 & 0 & 0 & 0 & 0 & 0 & 0 & 0 & 0 & 0 & 0 & 0 & 0 & 0 & 0 & 0 & 0 & 0 & \\ 
0 & 0 & 0 & 0 & 0 & 0 & 0 & 0 & 0 & 0 & 0 & 0 & 0 & 0 & 0 & 0 & 1 & 0 & 0 & 0 & 0 & 0 & \\
\end{bmatrix}
\nonumber
\end{equation}

\subsection{Dati di input}
I dati di input fondamentali per l'algoritmo sono le quantità di energia in vendita ed in acquisto ed i relativi prezzi.
Per ogni offerta è specificato:
\begin{itemize}
\item	CD\_PURPOSE: specifica se l'offerta presentata è una vendita OFF o un acquisto BID. Nel caso sia OFF viene mappato con $1$ altrimenti con $-1$.
\item N\_INTERVAL: l'ora a cui fa riferimento l'offerta di vendita o acquisto di energia.
\item CD\_ZONE: la zona a cui appartiene l'offerta. Ogni zona è mappata con un indice numerico $i=1,\dots,N$.
\item N\_QUANTITY: la quantità di offerta presentata in $MWh$.
\item N\_ENERGY\_PRICE: il prezzo a cui viene presentata l'offerta.
\end{itemize}

Altro input necessario per la corretta risoluzione del problema sono i limiti di transito tra zone, ovvero le quantitià di energia massime che possono transitare sui connettori. Per ogni limite viene specificato:
\begin{itemize}
\item DA: zona di partenza del connettore, viene mappata in un id numerico coerente con quello usati per mappare CD\_ZONE, con $i=1,\dots,N$.
\item A: zona di arrivo del connettore, viene mappata in un id numerico coerente con quello usati per mappare CD\_ZONE, con $j=1,\dots,N$.
\item LIMITE\_TRANSITO: massima quantità di energia che può transitare sul connettore, viene indicato con $MAX_{ij}$.
\end{itemize}

A fronte di questi dati di input il problema è ben definito e può essere implementato numericamente e risolto tramite un qualsiasi software di ottimizzazione lineare.

\subsection{Implementazione e risoluzione numerica}

L'algoritmo FIEMS. viene eseguito su ogni ora per la quale si vuole calcolare il prezzo dell'energia ed è riassunto nel Listato $\eqref{FIEMS_listato}$

\begin{algorithm}
\caption{FIEMS}
\label{FIEMS_listato}
\begin{algorithmic}[1]
\STATE Filtra le offerte di energia in vendita e in acquisto per ora di flusso. \label{FIEMSis1}
\STATE Costruisci la funzione obiettivo. \label{FIEMSis2}
\STATE Scrivi i vincoli sui transiti ammissibili tra zone nella forma $Ax\le b$. \label{FIEMSis3} ed il vincolo sul bilancio di energia.
\STATE Imponi i vincoli sulle variabili decisionali $QA_{k_c} \le QOA_{k_c} \; \forall \; k_c\in{1,2,\dots,KC}$ e $QV_{k_g} \le QOV_{k_g} \; \forall \; k_g\in{1,2,\dots,KG}$. \label{FIEMSis4}
\STATE Risolvi il problema di ottimizzazione lineare così composto ottenendo i valori delle variabili $QV_{k_g}$ e $Q_A{k_c}$. \label{FIEMSis5}
\STATE Verifica quali zone si sono separate e per ogni macrozona formatasi determina il prezzo dell'energia in quella zona. \label{FIEMSis6}
\STATE Calcola i transiti tra zone. \label{FIEMSis7}
\end{algorithmic}
\end{algorithm}
I punti $\eqref{FIEMSis1},\eqref{FIEMSis2}$ sono facilmente implementabili in $Matlab$. I filtri sono facilmente eseguibili per mezzo della funzione $Matlab$ find mentre la scrittura della funzione obiettivo, essendo lineare è scrivibile sfruttando i prodotti vettoriali tra i vari vettori interessati. Definito $\boldsymbol{Pz}$ il vettore dei prezzi e $\boldsymbol{T}$ il vettore dei tipi di offerta dove $T_{i}=1$ se l'offerta $i$-esima è una vendita e $T_{i}=-1$ se l'offerta $i$-esima è un acquisto la funzione obiettivo $F_{obj}$ può essere scritta come 
\begin{equation}
F_{obj}=\left \langle \boldsymbol{Pz}, \boldsymbol{T} \right \rangle,
\nonumber
\end{equation}
dove $\left \langle \cdot,\cdot \right \rangle$ denota l'usuale prodotto scalare in $\mathbb{R}^n$. 
\par Una maggiore complessità, invece, si presenta per la scrittura dei vincoli sui limiti di transito. Il solver di $Matlab$ linprog, necessita che i vincoli di diseguaglianza lineari siano scritti nella forma $Ax\le b$, dove $A$ è una matrice,$x$ il vettore delle variabli decisionali e $b$ in termine noto. Bisogna avere quindi cura di scrivere i vincoli sui transiti in questa precisa forma.\\ Il calcolo del limite di transito sul generico connettore $i,j$ può essere scritto come differenza tra la somma delle quantità in vendita e la somma delle quantità in acquisto nelle zone connesse ad $i$ aprendo l'arco $i,j$. Determinare quali siano le zone connesse ad una determinata zona richiede un algoritmo di ricerca sui grafi: nel nostro caso abbiamo usato un algoritmo di tipo \textit{Depth-first search}, implementato appoggiandosi sulle stack di Java disponibili utilizzando $Matlab$: esistono numerosi algoritmi di questo tipo in letteratura: il più semplice è riportato nel Listato $\eqref{DEPTHFIRST_LISTATO}$

\begin{algorithm}
\caption{Depth-fist search}
\label{DEPTHFIRST_LISTATO}
\begin{algorithmic}[1]
\STATE Initialize an empty stack, $S=stack()$. 
\STATE Initialize a boolean vector of $N$ elements to false, $V\left[1,\dots,N\right]=false$.
\STATE Open edge $i,j$, $G\left(i,j\right)=0$.
\STATE Select starting node $i$ and $S.push(i)$. \label{DEPTHFIRST_LISTATO_start_node}
\WHILE {$isnotempty(S)$}
\STATE {$n=S.pop()$}
\FOR{$k=1$ to $N$}
\IF{$G(n,k)=1$ and $V(k)=false$ }
\STATE $S.push(k)$
\STATE $V(k)=true$
\ENDIF
\ENDFOR
\ENDWHILE
\end{algorithmic}
\end{algorithm}

Grazie a tale algoritmo è possibile selezionare tutte le offerte di acquisto e vendita di energia che concorrono a formare il transito sull'arco di interesse e scrivere il vincolo nella forma desiderata $Ax\le b$ dove, ad esempio, 
\begin{equation}
A=
\begin{bmatrix}
	 1		& -1     & 0      & 1      & 1	    & \cdots & 1	  & 0     \\
	 \vdots & \vdots & \vdots & \vdots & \vdots & \vdots & \vdots & \vdots \\
	 -1		& -1     & 1      & 0      & 0	    & \cdots & 1	  & -1     \\
\end{bmatrix}
\nonumber
\end{equation}

\begin{equation}
x=
	\begin{bmatrix}
		 x_1\\
		 x_2\\
		 \vdots\\
		 x_{KG+KC}
   \end{bmatrix}
   \nonumber
\end{equation}
e 
\begin{equation}
b=
\begin{bmatrix}
10000\\
2000\\
\vdots\\
800
\end{bmatrix}
\nonumber
\end{equation}

Il vincolo sull'equilibrio del sistema è invece di facile scrittura ed essendo un vincolo di eguaglianza dovrà essere scritto in $MATLAB$ sfruttando la scrittura $A_{eq}x=b_{eq}$ con $A_{eq}=T$ e $b_{eq}=0$. Tale vincolo sta ad indicare che nel sistema complessivo, l'energia consumata deve essere pari a quella prodotta.
\par I vincoli sulle variabili decisionali \( Punto $\eqref{FIEMSis4}$\), invece, sono stati scritti imponendo i parametri upper bound $ub$ e lower bound $lb$ di Matlab rispettivamente pari a $\left[QV,QA\right]$ e a $\boldsymbol{0}$. Facendo così non sarà possibile accettare un'offerta per una quantità di energia superiore a quella presentata o negativa.
\par Infine per eseguire il passo $\eqref{FIEMSis5}$ dell'algoritmo è stato chiamato il risolutore linprog passandogli come parametri i valori $F_{obj}$,$A$,$b$,$Aeq$,$beq$,$lb$,$ub$ sopra definiti. In output il risolutore restituisce il vettore $x$ che contiene la parte di offerta accettata per ogni singola offerta presentata
\footnote{\label{note_prezzi}La particolare scrittura del problema dovrebbe portare l'ottimizzatore ad accettare tutte offerte che vengono accettate per una quantità pari a quella offerta e a rifiutare per intero le altre ed ad ammettere al più una sola offerta parzializzata nella singola macrozona. In realtà le offerte che vengono rigettate non presentano una $x_i=0$ ma dell'ordine di $10^{-8}$. Per ovviare a questo problema tutte le $x_i\le 10^{-4}$ vengono forzate a $0$. Inoltre se le ultime offerte di vendita accettate hanno prezzi molto simili tra loro l'ottimizzatore tende a parzializzarle entrambe. Per risolvere questo problema per ogni macrozona si fa il calcolo della quantità di totale energia in vendita accettata $QV_{z}$ e si vanno a modificare le $x_i$ delle offerte di vendita ordinate per prezzo appartenenti a quella macrozona ponendo $x_i=QOV_{i}$ fino al raggiungimento di $QV_{z}$.} ed i moltiplicatori di Lagrange $\boldsymbol{\mu}$ associati ai vincoli di diseguaglianza e il vettore $\boldsymbol{\lambda}$ associato ai vincoli di eguaglianza. Questi ultimi saranno fondamentali per determinare le macro zone formatesi. 

\subsection{Determinazione delle macrozone e dei prezzi zonali}
\begin{figure}[!h]
	\centering
	\includegraphics[scale=0.8]{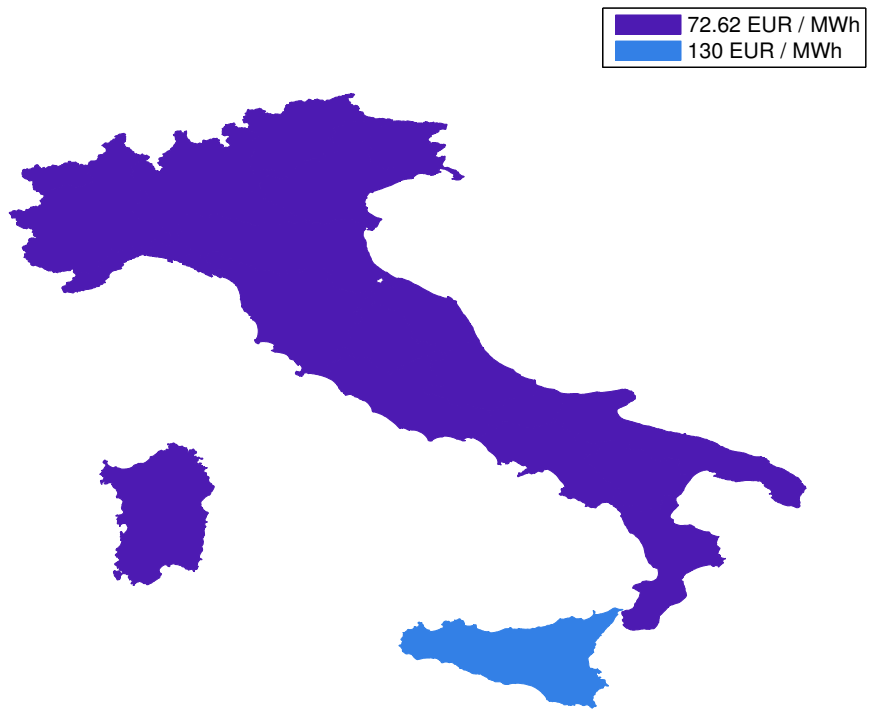}
	\caption{Esempio esito di mercato}
	\label{FIG_ESITO_MRK_ESEMPIO}
\end{figure}

Il passo successivo dell'algoritmo è quello di determinare se e in quali zone si è separato il mercato e, successivamente, quali sono i prezzi delle singole zone.\\ 
Per definizione si ha una separazione tra zone nel momento in cui un limite di transito si è saturato. Ad esempio in Figura $\eqref{FIG_ESITO_MRK_ESEMPIO}$ la Sicilia si è separata dal resto dell'Italia, con conseguente diversificazione dei prezzi, perchè si è saturato il limite di transito che la collega al continente.
Per determinare se c'è stata una separazione tra zone è sufficiente osservare i moltiplicatori di Lagrange $\boldsymbol{\mu}$ associati ai vincoli sui limiti di transito. Se il moltiplicatore di Lagrange $\mu_s$ associato al vincolo $s$-esimo sul limite di transito tra $i$ e $j$ è diverso da $0$ allora il connettore si è saturato. Ragionando così si possono determinare le varie macrozone formatesi secondo l'algoritmo riportato nel Listato $\eqref{MACROZONE_listato}$.

\begin{algorithm}
	\caption{Algoritmo di determinazione delle macrozone.}
	\label{MACROZONE_listato}
	\begin{algorithmic}[1]
		\STATE Open all edges $G_{ij}$ whose Lagrange multipliers $\mu_s\ne 0$.
		\STATE Inizialize an empty matrix $\boldsymbol{M}$ and a boolean vector $V$ such that $V\left[1,\dots,N\right]=false$.
		\STATE Set $k=1$.
		\WHILE  {$V\left(i\right)=true$ for all $i \in N$}
			\STATE Select node $n$ such that $V\left[i\right]=false$.
			\STATE Run Algorithm $\eqref{DEPTHFIRST_LISTATO}$ starting from row $\eqref{DEPTHFIRST_LISTATO_start_node}$ with start node equal to $n$. You get a vector $V_{p}$ which is a boolean vector such that $V_{p}\left[i\right]=1$ if node $i$ has been visited, $0$ otherwise. \label{MACROZONE_listato_1}
			\STATE Define a vector $Y$ which contains visited nodes by run $\eqref{MACROZONE_listato_1}$.
			\STATE  Set $V=V+V_{p}$.
			\STATE Set $\boldsymbol{M}\left(k,:\right)=Y$.
		\ENDWHILE
		\STATE Return $\boldsymbol{M}$.
	\end{algorithmic}
\end{algorithm}

Al termine dell'Algoritmo $\eqref{MACROZONE_listato}$ ogni riga della matrice $\boldsymbol{M}$ conterrà gli indici numerici corrispondenti alle zone che appartengono alla stessa macrozona. Il procedimento di determinazione delle macrozone è così concluso.
\par E' possibile ora passare alla determinazione dei prezzi di mercato. Seguendo quanto descritto in precedenza il prezzo della zona $\rho_z$ è dato da 

\begin{equation}
\rho_z=\lambda-\sum_{ij \; s.t. \; C_{ij}=1} \mu_{ij}S_{ij}^z - \sum_{\alpha=1,2,\dots,M}\nu_{\alpha} A_{\alpha}^z.
\nonumber
\end{equation}
che nel caso la rete sia ad albero assume la forma semplificata seguente:
\begin{equation}
\rho_z=\lambda-\sum_{ij \; s.t. \; C_{ij}=1} \mu_{ij}.
\label{Prezzo_analitico}
\end{equation}
Tutti gli elementi dell'equazione $\eqref{Prezzo_analitico}$ sono noti ed il prezzo per ogni zona risulta essere così calcolabile\footnote{Un problema nell'ultilizzo di questo metodo sorge nel momento in cui non è noto il dettaglio delle offerte ed in cui si vuole testare l'algoritmo con il mercato reale in cui invece è noto l'intero dettaglio delle offerte per le zone estere. Quindi per le zone BSP, XFRA, XAUS e MALT sono note solo le quantità di vendita e acquisto aggregate accettate. Non è quindi possibile testare perfettamente l'output dell'algoritmo con l'esito di mercato reale in quanto i moltiplicatori di Lagrange $\mu_s$ non sono calcolati correttamente. In fase di modellizzazione si assume quindi che le offerte di vendita nelle zone estere siano poste a $0$ mentre quelle di acquisto senza indicazione di prezzo. Il prezzo zonale allora viene calcolato come il prezzo corrispondente all'ultima offerta in ordine di prezzo crescente di vendita accettata nella macrozona (dove il criterio di accettazione dell'offerta è quello specificato nalla Nota \ref{note_prezzi}). Tale approccio, sebbene ancora non permetta un confronto perfetto con gli esiti reali di mercato in quanto l'algoritmo di mercato è ben più complesso e coinvolge anche il calcolo del PUN, ha permesso di ridurre gli errori commessi dal nostro algoritmo.}.
\par Il calcolo dei transiti tra zone è estremamente semplice. Infatti basta notare nel vincolo $Ax \le b$ la posizione $k$ del vettore $Tr=Ax$ rappresenza il transito sull'arco $k$. Quindi i transiti tra zone sono calcolabili come $Tr=Ax$ \footnote{Nel nostro caso, i coeffincienti $S_{ij}$ e $A_{\alpha}^z$ non sono stati resi pubblici dal GME nel momento in cui l'anello CNOR-CORS-SARD-CSUD è stato chiuso. L'algoritmo continua a produrre dei prezzi verosimili, ma i transiti non possono più essere calcolati come $Tr=Ax$. Per ovviare parzialmente a questo problema viene aperto il transito CNOR-CORS che risulta essere il meno capiente e i transiti sugli altri archi dell'anello vengono ricalcolati secondo un bilancio sulle energie vendute-acquistate sulla singola zona.}.

\section{Risulati numerici}
In questa sezione analizzeremo brevemente le performance numeriche dell'algoritmo ed i risulati da esso prodotti. Gli esperimenti sono stati eseguiti sul un PC Desktop con processore Intel\(R\) Core\(TM\) i3-2100 CPU @ 3.10 GHz e RAM 4.00 GB.\\
\par
\begin{figure}[h]
	\centering
	\includegraphics[scale=0.7]{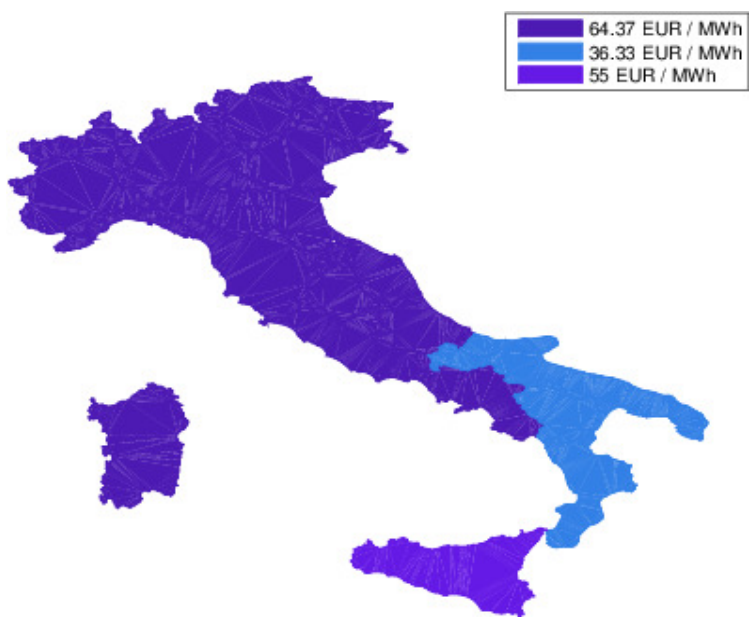}
	\caption{Esito di mercato ora 9 del 4 marzo 2014.}
	\label{italia_HF9}
\end{figure}

 Consideriamo per gli esperimenti il giorno 4 marzo 2014, limitatamente all'ora 9.
La situazione creatasi è mostrata in Figura $\ref{italia_HF9}$ .

Si nota come il mercato l'Italia si sia separata in tre macrozona a cui corrispondo tre differenti prezzi. \\
I prezzi reali e calcolati dall'algoritmo sono riportati in Tabella $\ref{Tabella__1}$.

\begin{table}[!h]
	\centering
	\begin{tabular}{ccc}
		\toprule
		$Zona$ &  $Prezzo reale$      & $Prezzo algoritmo$ \\
		$-$     & $EUR/MWh$   		  & $EUR/MWh$\\
		\midrule
		NORD & 64.37 & 64.37	\\ 
		CNOR & 64.37 & 64.37	\\ 
		CSUD & 64.37 & 64.37	\\ 
		SUD &  36.33 & 36.33	\\ 
		SICI & 55    & 55   	\\ 
		SARD & 64.37 & 64.37	\\
		BSP  & 60.02 & 0 \\
		\bottomrule
	\end{tabular}
	\caption{Prezzi zone Italiane.}
	\label{Tabella__1}
\end{table}

I risultati sono perfetti, tranne che per il BSP dove è evidente l'assunzione fatta sulle zone estere.\\
In Tabella $\ref{Tabella__2}$ sono confrontati i transiti sui connettori reali e quelli calcolati dall'algoritmo, limitatamente ai transiti non nulli.
\begin{table}[!h]
	\centering
	\begin{tabular}{ccccc}
		\toprule
		$Da$ & $A$ &  $Transito reale$      & $Transito calcolato$ & $Differenza$ \\
		$-$  & $-$ &  $MWh$  		  	    & $MWh$                & $MWh$\\
		\midrule
		SUD	&	ROSN	&	-750	&	-750	&	0,00 \\
		SLOV	&	BSP	&	-649	&	-649	&	0,00 \\
		CNOR	&	CSUD	&	-2519,103	&	-2519,103001	&	0,00 \\
		CSUD	&	SARD	&	-188,312	&	-188,312	&	0,00 \\
		CSUD	&	SUD	&	-2800	&	-2800	&	0,00 \\
		NORD	&	AUST	&	-280	&	-280	&	0,00 \\
		NORD	&	CNOR	&	-1039,883	&	-1039,883001	&	0,00 \\
		NORD	&	FRAN	&	-2785	&	-2785	&	0,00 \\
		NORD	&	SLOV	&	-649	&	-649	&	0,00 \\
		NORD	&	SVIZ	&	-3769	&	-3769	&	0,00 \\
		SARD	&	CORS	&	49	&	49	&	0,00 \\
		SICI	&	PRGP	&	-92,864	&	-104,0710455	&	11,21 \\
		SICI	&	ROSN	&	-100	&	-100	&	0,00 \\
		SUD	&	BRNN	&	-1710,001	&	-1749,842	&	39,84 \\
		SUD	&	FOGN	&	-323,765	&	-323,765	&	0,00 \\
		\bottomrule
	\end{tabular}
	\caption{Transiti tra zone.}
	\label{Tabella__2}
\end{table}

Osserviamo che questi risultati sono molto buoni, anche se va segnalato che in alcuni casi si possono manifestare differenze di prezzo in una macrozona dell'ordine di $1.5 \, EUR/MWh$. Talvolta, sopratutto in nella zona SICI, le differenze possono essere più marcate, anche dell'ordine dei $10/20 \, EUR/MWh$. Tale discrepanza può essere spiegata sia per il fatto che l'algoritmo non gestisce situazioni strane di incrocio tra domanda e offerta come quelle descritte in . Inoltre, il fatto che il PUN non venga calcolato a all'interno dell'algoritmo può generare questo tipo di errori. Nonostante ciò, l'eventualità di discrepanze nel prezzo di quantità superiori a  $1.5 \, EUR$ è molto bassa e l'algoritmo produce dei risultati plausibili nella maggior parte dei casi.\\
Da ultimo, una nota sul tempo computazionale. L'algoritmo di minimizzazione usato da Matlab per risolvere tale problema risulta essere estremamente efficiente: per determinare prezzi e transiti di un'ora di mercato il tempo medio calcolato su ventiquattro simulazioni è di $0.24$ secondi. Tale velocità di esecuzione unita ai risultati con errori accettabili prodotti dall'algoritmo del GME semplificato, non rende necessaria un'implementazione più dettagliata, come quella descritta in. Difatti, la risoluzione, ad esempio, di un problema di programmazione lineare parametrico (necessario per la determinazione corretta del PUN) comporterebbe un'aumento del tempo di esecuzione ed aggiungerebbe poca sostanza alle analisi per il quale questo sviluppo è stato richiesto.

\bibliographystyle{plain}
\clearpage
\bibliography{Bibliografia}

\end{document}